\documentclass[aps,prl,twocolumn,amsmath,amssymb,showpacs]{revtex4-1}

\usepackage{graphicx}
\usepackage{dcolumn}
\usepackage{bm}
\usepackage{xcolor}
\usepackage{siunitx}
\usepackage{hyperref}
\usepackage{amsmath}
\usepackage{amssymb}
\usepackage{sidecap}
\usepackage{mathtools}
\usepackage{float}
\usepackage{physics}
\usepackage{booktabs}

\newcommand{\gstate}{$^4$I$_{15/2}$~}
\newcommand{\estate}{$^4$I$_{13/2}$~}

\begin{document}

\title{Characterization of the spin and crystal field Hamiltonian of erbium dopants in silicon}

\author{Adrian Holzäpfel$^{1,2}$}
\altaffiliation[Contributed ]{equally to this work}
\author{Stephan Rinner$^{1,2}$}
\altaffiliation[Contributed ]{equally to this work}
\author{Kilian Sandholzer$^{1,2}$}
\author{Andreas Gritsch$^{1,2}$}
\author{Thierry Chanelière$^{3}$}
\author{Andreas Reiserer$^{1,2}$}
\email{andreas.reiserer@tum.de} 

\affiliation{$^{1}$Max-Planck-Institute of Quantum Optics, Quantum Networks Group, Hans-Kopfermann-Stra{\ss}e 1, D-85748 Garching, Germany \\
$^{2}$Technical University of Munich, TUM School of Natural Sciences and Munich Center for Quantum Science and Technology (MCQST), James-Franck-Stra{\ss}e 1, D-85748 Garching, Germany\\
$^{3}$Univ. Grenoble Alpes, CNRS, Grenoble INP, Institut Néeel, 38000 Grenoble, France
}

\date{\today}

\begin{abstract}
The integration of coherent emitters into low-loss photonic circuits is a key technology for quantum networking. In this context, nanophotonic silicon devices implanted with erbium are a promising hardware platform that combines advanced wafer-scale nanofabrication technology with coherent emission in the minimal-loss band of optical fibers. Recent studies have reported two distinct sites in the silicon lattice in which erbium can be reproducibly integrated with particularly promising properties. Here, for an in-depth analysis of these sites, resonant fluorescence spectroscopy is performed on a nanophotonic waveguide in magnetic fields applied along different orientations. In this way, the site symmetry is determined, the spin Hamiltonian is reconstructed and a partial fit of the crystal field Hamiltonian is performed. The obtained quantitative description of the magnetic interaction allows the optimization of Zeeman splittings, optical branching ratios or microwave driving to the needs of future experiments. Beyond that, the derived site symmetry constrains the location of the erbium dopant in the silicon unit cell. This is a key step towards a detailed microscopic understanding of the erbium sites, which may help to improve the integration yield, thus paving the way to efficient nanophotonic quantum memories based on the Er:Si platform.
\end{abstract}

\maketitle

\section{Introduction}
Quantum information processing based on optically addressable solid-state spin qubits has attracted growing interest in recent years.\textsuperscript{\cite{awschalom_quantum_2018}} Pioneering experiments have investigated color centers in diamond\textsuperscript{\cite{{bernien_heralded_2013, sipahigil_integrated_2016, parker_diamond_2024}}} or single rare-earth dopants in yttrium-based crystals\textsuperscript{\cite{kindem_control_2020, chen_parallel_2020, ulanowski_spectral_2022}} and calcium tungstate.\textsuperscript{\cite{ourari_indistinguishable_2023, uysal_spin-photon_2024}} However, since these crystals are not compatible with wafer-scale nanofabrication technology, their use in photonic integrated circuits requires complex heterogeneous integration schemes.\textsuperscript{\cite{chen_parallel_2020, li_heterogeneous_2024}} This can be avoided by using color centers\textsuperscript{\cite{bergeron_silicon-integrated_2020, redjem_single_2020}} or erbium dopants\textsuperscript{\cite{weiss_erbium_2021, gritsch_narrow_2022, berkman_millisecond_2023}} in silicon. Corresponding systems can be fabricated using wafer-scale processes\textsuperscript{\cite{rinner_erbium_2023}} and thus offer a high potential for up-scaling. \\
\begin{sloppypar}
This work focuses on erbium dopants in their commonly found triply ionized charge state, $\mathrm{Er}^{3+}$. 
With a wavelength of $\approx\SI{1.53}{\micro\meter}$, their optical transitions between the $^4\mathrm{I}_{15/2}$ ground state and the $^4\mathrm{I}_{13/2}$ state falls within the minimum loss band of optical fibers, and can exhibit outstanding coherence properties both in bulk crystals\textsuperscript{\cite{bottger_optical_2006, ulanowski_spectral_2022}} and in photonic nanostructures.\textsuperscript{\cite{gritsch_narrow_2022, ourari_indistinguishable_2023}} 
Together with their long spin-state coherence,\textsuperscript{\cite{rancic_coherence_2018, wang_single-electron_2023, berkman_millisecond_2023}} this makes erbium dopants a promising hardware platform for quantum networking.
\end{sloppypar}

However, it is not straightforward to reliably integrate erbium dopants into silicon because of the ionic size mismatch and the different bond types, which results in a low solubility.\textsuperscript{\cite{kenyon_erbium_2005}} Thus, non-equilibrium methods such as ion implantation, or integration during growth via molecular beam epitaxy or chemical vapor deposition are required. In early work, it has been observed that erbium is clustering and gettering in silicon, i.e., it easily bonds to other impurities and lattice defects, which result in a large number of possible integration configurations that depend on the chemical purity of the silicon host and the implantation and annealing conditions.\textsuperscript{\cite{kenyon_erbium_2005, weiss_erbium_2021, berkman_observing_2023}} Eventually, reproducible integration of erbium dopants in silicon at a small number of well-defined lattice sites with a narrow inhomogenenous linewidth has only been demonstrated recently,\textsuperscript{\cite{gritsch_narrow_2022, rinner_erbium_2023}} paving the way to single-dopant spectroscopy\textsuperscript{\cite{gritsch_purcell_2023}} and optical spin readout.\textsuperscript{\cite{gritsch_optical_2024}} 

While the optical properties of these sites termed "A" and "B" have been characterized, their magnetic interaction has not been investigated in detail. This is the focus of this work for which we use optical fluorescence spectroscopy in a magnetic field applied along different directions. Based on these measurements, the symmetry of the integration sites is characterized, and an effective spin Hamiltonian is determined for site A. This is then compared to the prediction from a partial fit of a crystal field (CF) Hamiltonian restricted to the measured CF structure of the $J=15/2$ and $J=13/2$ manifolds.

\section{Spin Hamiltonian}\label{SEC:spinHamil}

\begin{figure*}[t]
  \includegraphics[width=\linewidth]{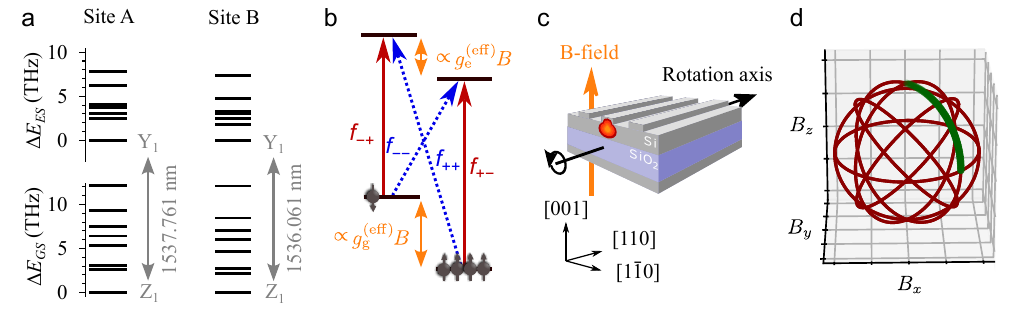}
  \caption{(a) The interaction of the erbium dopants with the surrounding crystal, described with a crystal-field Hamiltonian, lifts the degeneracy of both the $^{4}\mathrm{I}_{15/2}$  multiplet "Z" and the $^{4}\mathrm{I}_{13/2}$ optically excited multiplet "Y" for sites with sufficiently low point symmetry --- here shown as level schemes for the crystallographic sites A and B. The numeric values of the detunings can be found in Table~\ref{TAB:CF}. (b) By applying an external magnetic field, the remaining degeneracy of the Kramers doublets is lifted, as described by a spin Hamiltonian. Sufficiently low temperatures and high magnetic fields lead to a population imbalance (spin symbols) between the two ground states so that the two spin-flip ($f_\mathrm{--}$, $f_\mathrm{++}$) and two spin-preserving ($f_\mathrm{+-}$, $f_\mathrm{-+}$) transitions can be unambiguously identified. (c) Experimental setting. Erbium dopants are integrated in the silicon device layer of a silicon-on-insulator chip and couple to the guided optical mode (red) of a nanophotonic rib waveguide. To determine the g-tensor, fluorescence spectra are recorded while the sample is rotated around in the [110] crystal axis in an external magnetic field. (d) Effective bias field directions on the unit sphere. Fluorescence spectra have only been recorded in a $90^\circ$ range. Because of the high symmetry of the silicon host crystal, this is fully sufficient to determine the g-tensor, as the dataset contains all information that can be gained by probing along any axis (red) that is related to the measurement directions (green) by symmetry operations of the host (see section~\ref{SEC:sym}).}
  \label{fig1}
\end{figure*}

For both, site A and site B, the degeneracy of the multiplets within the 4f electronic orbitals is fully lifted by the crystal field.\textsuperscript{\cite{gritsch_narrow_2022}} Figure~\ref{fig1}~(a) shows the resulting eight crystal fields levels for the ground state and seven for the excited state for each site. As erbium is a Kramers dopant, each crystal field level is doubly degenerate. Each of these Kramers' doublets can be described by an effective electron spin $S=1/2$ with an anisotropic Zeeman coupling using a Spin Hamiltonian $\mathcal{H}_\mathrm{spin}$ of the form:
\begin{equation}
    \mathcal{H}_\mathrm{spin} = \mu_B\cdot\vec{B}\cdot\mathbf{g}\cdot\vec{S}.
\end{equation}

Here, $\mu_B$ is the Bohr magneton, $\vec{B}$ is the externally applied magnetic field and $\vec{S}\coloneqq  \frac{1}{2} (\hat{\sigma}_x,\hat{\sigma}_y,\hat{\sigma}_z)$ is a vector containing the three Pauli operators. The interaction is parameterized by the g-tensor $\mathbf{g}$, a $3\times 3$ symmetric matrix. It is convenient to define the effective g-factor $g^\mathrm{(eff)}$ which quantifies the magnitude of the Zeeman effect for a given field direction:
\begin{equation}
    g^\mathrm{(eff)} = \frac{|\vec{B}\cdot\mathbf{g}|}{|\vec{B}|}.
\end{equation}

This effective tensor generally differs between the $Z_1$ ground state $(g^\mathrm{(eff)}_\mathrm{g})$ and the $Y_1$ optically excited state $(g^\mathrm{(eff)}_\mathrm{e})$. As illustrated in Figure~\ref{fig1}~b, four optical transitions can thus be observed in spectroscopy: Two spin-preserving ones at frequencies $f_\mathrm{+-}$ and $f_\mathrm{-+}$, as well as two spin-flip transitions at frequencies $f_\mathrm{--}$ and $f_\mathrm{++}$:
\begin{equation}\label{EQU:branches}
    \begin{split}
        f_\mathrm{++}&=f_\mathrm{0}+\frac{\mu_B}{2h}(g^\mathrm{(eff)}_\mathrm{g}+g^\mathrm{(eff)}_\mathrm{e})B,\\
        f_\mathrm{+-}&=f_\mathrm{0}+\frac{\mu_B}{2h}(g^\mathrm{(eff)}_\mathrm{g}-g^\mathrm{(eff)}_\mathrm{e})B,\\
        f_\mathrm{-+}&=f_\mathrm{0}+\frac{\mu_B}{2h}(-g^\mathrm{(eff)}_\mathrm{g}+g^\mathrm{(eff)}_\mathrm{e})B,\\
        f_\mathrm{--}&=f_\mathrm{0}+\frac{\mu_B}{2h}(-g^\mathrm{(eff)}_\mathrm{g}-g^\mathrm{(eff)}_\mathrm{e})B.\\
    \end{split}
\end{equation}
Here, $f_0$ is the transition frequency when no bias field is applied, $B$ is the field magnitude and $h$ is Planck's constant. The correct assignment of the spin-preserving transitions $f_\mathrm{+-}$ and $f_\mathrm{-+}$ within an observed spectrum may be ambiguous because it is a priori not clear if $g_g^{(\mathrm{eff})} > g_e^{(\mathrm{eff})}$ or vice versa. This can be resolved by noting that at low temperatures and high fields the lowest energy state in the system will be the most populated and, consequently, fluorescence at $f_\mathrm{+-}$ will be brighter than fluorescence at $f_\mathrm{-+}$ (Figure~\ref{fig1}~(b)). 

The spin Hamiltonians of the ground and excited state can be determined by finding the g-tensors $\mathbf{g}_\mathrm{g}$ and $\mathbf{g}_\mathrm{e}$ that reproduce the transition frequencies in Equation~\ref{EQU:branches} for all possible field directions. Depending on the crystal symmetry, this may require measurements of the magnetic interaction with the field rotating in up to three independent planes.\textsuperscript{\cite{weil_electron_2007}} In silicon, only one plane is sufficient, as will be detailed in the following.

\subsection{Subsites and symmetry}\label{SEC:sym}
Crystalline silicon is organized in a diamond cubic structure and, as such, has full octahedral symmetry $(\mathrm{O_h})$. Whenever a defect is introduced into the crystal structure, the resulting site may preserve some of the point symmetries of the host material while breaking others. When a symmetry is broken by the site, then applying the corresponding symmetry operation maps the site onto a different orientation, i.e., to a different \emph{subsite}. Different subsites have identical properties and degenerate transitions unless an additional interaction such as strain, Stark or Zeeman effect introduces a preferred direction and breaks the degeneracy.

The number of subsites $k$ follows the relation: 
\begin{equation}
    N_{CS}=N_{S} \times k, \label{EQU:N_sites}
\end{equation} 
where $N_{CS}$ is the order of the point group of the crystal and $N_{S}$ is the order of the point group of the integration site of the defect.\textsuperscript{\cite{spaeth_structural_2012}} For the octahedral symmetry of the silicon crystal, this means that in general, up to 48 subsites are possible for a crystallographic site. Because of the inversion symmetry of the Zeeman interaction, this reduces to at most 24 classes of magnetically inequivalent subsites. 

As the subsites are related by crystal symmetries, knowing the tensor $\mathbf{g}_0$ of an arbitrary magnetic class, the $\mathbf{g}_i$ of all magnetic classes can be constructed  by applying all symmetry operations $\mathbf{\Pi_i}$ from the point group of the host material:
\begin{equation}\label{EQU:syms}
    \mathbf{g}_i=\mathbf{\Pi}_i^{-1} \cdot \mathbf{g}_0 \cdot \mathbf{\Pi}_i.
\end{equation}
Instead of applying all 48 elements of $\mathrm{O_h}$, it is convenient to always just consider a single representative of classes of operations that only differ by inversion, as every operation within such a class has the same effect on the g-tensor. Therefore, it is sufficient to just consider the 24 elements of $\mathrm{T_d}$, i.e. the biggest subgroup of $\mathrm{O_h}$ that does not contain inversion.

In silicon and other materials with high symmetries, the relationship from Equation~\ref{EQU:syms} reduces the required number of measurements to determine $\mathbf{g}_0$ unambiguously. Whenever the Zeeman interaction is probed along a direction $\left[hkl\right]$, this measurement contains all of the information that could be gained by probing the crystal along the set of directions $\langle hkl\rangle$ that are related by crystal symmetries. So, while in crystals with low symmetry one needs to record the Zeeman effect in three different planes to fully determine the g-tensor,\textsuperscript{\cite{weil_electron_2007}} any single plane is sufficient in silicon, which reduces the required effort in the following measurements.

\subsection{Results}

\begin{figure*} 
     \includegraphics[width=\linewidth]{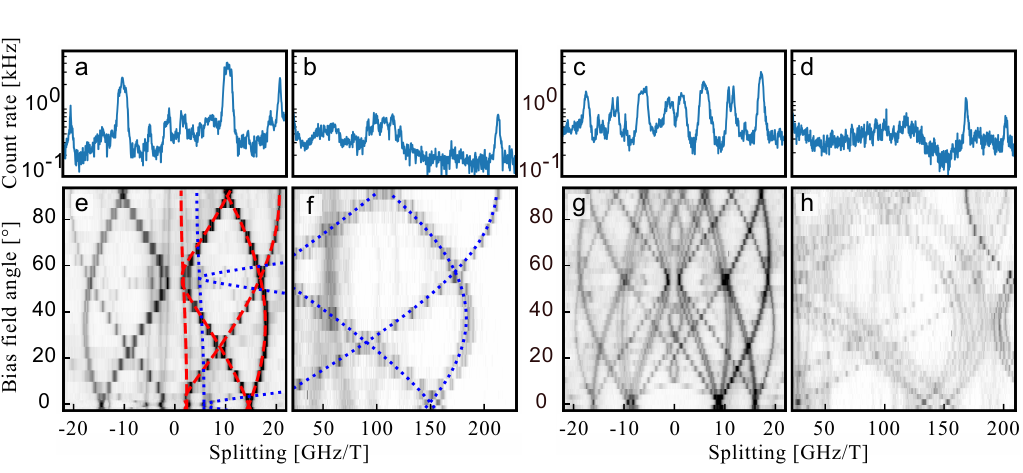}
        \caption{Pulsed, resonant fluorescence spectra are recorded for different angles of the magnetic field when rotating the sample around the $[110]$ crystal axis in an external bias field. Panels (a-d) show example spectra recorded with the magnetic bias field along $[1\Bar{1}0]$ for site A (a,b) and site B (c,d). Panels (e-h) show the dependence on the field direction for the same spectra. The sample is rotated such that the field angle varies from $[001]$ to $[1\Bar{1}0]$, i.e., the spectra (a-d) correspond to a linecut at $90^\circ$ in (e-h). The different spectra are recorded at different bias field magnitudes to reduce their overlap with fluorescent background from other defects in the sample and resolve the respective feature to our best ability, with (a,c,e,g) at $B$=\SI{1.9}{\tesla}, (b,f) at $B$=\SI{0.25}{\tesla} and (d,h) at $B$=\SI{0.55}{\tesla}. In panel (e,f), we superimpose our fit to an effective spin Hamiltonian with $\mathrm{C_{2v}}$ symmetry, as presented in Equation~\ref{EQU:spinFit}. We indicate $f_{+-}$ (red dashes lines) and $f_{++}$ (blue dotted lines) as defined in Equation~\ref{EQU:branches} and omit the symmetric lines $f_{-+}$ and $f_{--}$ for a better visibility of the raw data. We note additional, slight non-degeneracies in the data, which can be seen as double lines closest to \SI{0}{\giga\hertz\per\tesla} in panel (e), and at the largest splittings in panel (f). Thus, the emitters exhibit a slight deviation from the imposed $\mathrm{C_{2v}}$ symmetry. A fit without this constraint is shown in Figure~\ref{FIG:genfit}.}
        \label{fig2}
\end{figure*}

To determine the Zeeman Hamiltonian, we apply an external bias field using a superconducting solenoid magnet. Then, based on the techniques introduced in \textsuperscript{\cite{gritsch_narrow_2022}}, we record pulsed resonant fluorescence spectra around the zero-field transition frequency for the two sites, A and B. Details of the setup and the measurement scheme can be found in part~\ref{subsec:setup}. We repeat this measurement while rotating the field direction. Based on the symmetry arguments outlined in section~\ref{SEC:sym}, we choose to measure only within the interval $0^\circ$ to $90^\circ$. This interval is non-redundant and at the same time contains all information that can be gained from $360^\circ$ rotation measurements around six different axes, as illustrated in Figure~\ref{fig1}~(d). Thus, the dataset shown in Figure~\ref{fig2} is sufficient for determining the effective spin Hamiltonian unambiguously.

\subsubsection{Site A}
For further analysis, the frequencies of the fluorescence peaks belonging to the site of interest are identified, as detailed in section~\ref{SEC:data_preprocessing}. Then, the spin Hamiltonian is fitted using a basin hopping algorithm that minimizes the distance of each data point to the closest model line. The details of this fitting procedure are described in section~\ref{SEC:Fitting}. From this fit, the following g-tensors are extracted:
\begin{equation}
    \mathbf{g}_\mathrm{g}=\begin{pmatrix}
  7.99 & 8.11 & 0.28\\
  8.11 & 8.71 & 0.11\\
  0.28 & 0.11 & 0.52\\
\end{pmatrix},~
  \mathbf{g}_\mathrm{e}=\begin{pmatrix}
  6.57 & 6.79 & 0.18\\
  6.79 & 7.07 & 0.12\\
  0.18 & 0.12 & 0.1\\
\end{pmatrix}\label{EQU:genfit}
\end{equation}
 For the average relative deviation (root mean square deviation, RMSD) of the fit we find $\Delta f/f\approx 4\%$. This estimate for the inaccuracy of the peak extraction and fitting procedure exceeds the systematic inaccuracy of the laser frequency measurement, the rotation angle and the magnetic field magnitude.

\begin{figure*}
     \includegraphics[width=\linewidth]{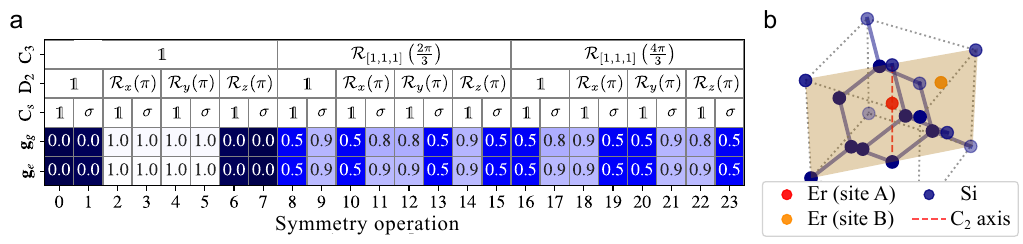}
        \caption{(a) Normalized commutators (as defined in Equation \ref{eq:commutators}) of the two fitted g-tensors with the elements of the point group of the host material. The symmetry for a given column can be calculated by concatenating the respective operations from the first three rows. (b) Unit cell of silicon with exemplary erbium positions for site A and site B based on their determined point symmetry. For site A with point symmetry $\mathrm{C_{2v}}$, the erbium dopant is located on the two-fold rotation axis that passes through the center of the unit cell along $\langle 100\rangle$. For site B with point symmetry $\mathrm{C_{s}}$, the erbium needs to be located in the $\{110\}$ mirror plane. Both sites may comprise additional impurity atoms or vacancies as long as their arrangement obeys the observed symmetries. }
  \label{fig3}
\end{figure*}

We now focus on the symmetry of the fitted g-tensors, which in turn implies the symmetry of the crystallographic site. If the $\mathbf{g}$-tensor exhibits a certain symmetry, its commutator with the corresponding symmetry operation $\mathbf{\Pi}_i$ will vanish. In the presence of any random or systematic error, however, the g-tensor may not respect any exact crystal symmetry. 
Thus, we determine how strongly a given symmetry $\mathbf{\Pi}_i$ is broken by calculating its normalized commutators with the g-tensor $\mathbf{g}$:
\begin{equation}  \label{eq:commutators}
    \mathcal{C}(\mathbf{g},\mathbf{\Pi}_i)=\frac{\left\lVert \mathbf{g}\mathbf{\Pi}_i - \mathbf{\Pi}_i\mathbf{g} \right\rVert_\mathrm{op} }{\left\lVert\mathbf{g}\right\rVert_\mathrm{op}}
\end{equation}
where $\left\lVert\cdot\right\rVert_\mathrm{op}$ is the operator norm. We summarize the commutators for both g-tensors with all symmetry operations from $\mathrm{T_d}$ in Figure~\ref{fig3}(a). For clarity, we choose to express each symmetry operation $\mathbf{\Pi}_i$ as a product of one element from each $\mathrm{C_3}$, $\mathrm{D_2}$ and $\mathrm{C_s}$, where $\mathrm{C_3}$ corresponds to three-fold rotation symmetry around $\left[111\right]$, $\mathrm{D_2}$ to two-fold rotation symmetry around the three crystal axes and $\mathrm{C_s}$ to mirror symmetry with regard to the $(110)$ plane. We find small commutators $\mathcal{C}(\mathbf{g},\mathbf{\Pi}_i) < 0.04$ for both g-tensors with the operations $\Pi_0,\Pi_1,\Pi_6$ and $\Pi_7$. These four operations represent the point group $\mathrm{C_{2v}}$ and we thus conclude that this is the point symmetry of site A. This symmetry corresponds to a $\{ 110\}$ mirror plane with an additional two-fold rotational axis along $\langle 001 \rangle$, lying within the mirror plane. The same symmetry has also been reported previously for oxygen-related Er-1 centers  in silicon.\textsuperscript{\cite{vinh_photonic_2009}} However, because of the strong difference of the CF levels, we can rule out that the two sites are identical.

Knowing the symmetry of the site allows us to locate the position of the erbium dopant in the silicon lattice, as illustrated in Figure~\ref{fig3}~(b). However, the microscopic environment of the site, i.e. whether the erbium is surrounded only by silicon atoms or is in fact a cluster of two or even several impurity atoms (as suggested for other sites \textsuperscript{\cite{vinh_photonic_2009, yang_zeeman_2022}}), remains an open question. Still, erbium integration in a substitutional or tetrahedral interstitial site would exhibit additional symmetries and can thus be ruled out. However, the symmetry of site A would be respected for an interstitial erbium atom at the center of the unit cell that features an additional displacement along the two-fold rotational axis, or a vacancy or substitutional defect atom on the top or bottom face of the fcc unit cell. Thus, these are plausible configurations for site A.\textsuperscript{\cite{hughes_crystal_2014}} However, further measurements such as nuclear spin spectroscopy\textsuperscript{\cite{abobeih_atomic-scale_2019}} will be required to determine the microscopic nature of the sites. 

After identifying the site symmetry, we repeat the fitting of the g-tensors while enforcing the $\mathrm{C_{2v}}$ symmetry by fixing their principal axes. This gives:

\begin{equation}\label{EQU:spinFit}
    \mathbf{g}_\mathrm{g}^\mathrm{(sym)}=\begin{pmatrix}
  8.5 & 8.1 & 0\\
  8.1 & 8.5 & 0\\
  0 & 0 & 0.58\\
\end{pmatrix},~
  \mathbf{g}_\mathrm{e}^\mathrm{(sym)}=\begin{pmatrix}
  6.94 & 6.72 & 0\\
  6.72 & 6.94 & 0\\
  0 & 0 & 0.24\\
\end{pmatrix}.
\end{equation}

This constrained fit is plotted along with the data in Figure~\ref{fig2}(e,f). As can be seen, it accounts for almost all features observed in the fluorescence spectra, except for minor additional splittings. For the error (RMSD), we find $\Delta f/f\approx 11\%$. We quantify how much the g-tensors from the unrestricted fit differ from the symmetry-restricted ones by averaging the relative deviation of the effective g-factors over all possible field directions:
\begin{equation}
    \Delta g=\left\langle \frac{|g^\mathrm{(gen)}-g^\mathrm{(sym)}|}{\frac{1}{2}\left(g^\mathrm{(gen)}+g^\mathrm{(sym)}\right)}\right\rangle.
\end{equation}

We find moderate deviations of $\Delta g_\mathrm{g}=6\%$ and $\Delta g_\mathrm{e}=7\%$ for the ground and excited state, respectively. These are attributed to the small additional splittings which cannot be described by the symmetrized fit resulting in a systematic contribution to its uncertainty. Consistently, we find that the RMSD of the symmetrized fit is approximately equal to the sum of these deviations and the RMSD without a symmetry constraint. We also report the symmetrized fit, despite its higher uncertainty, as we consider it likely that the additional splittings are a not a fundamental property of the site; instead, they may be particular to our sample, caused either by strain or by measurement imperfections such as sample misalignment. Thus, we expect that the symmetrized g-tensor is a more accurate description of the general magnetic properties of erbium dopants in site A.

We can predict a number of characteristics of site A based on the derived point symmetry. As $\mathrm{C_{2v}}$ has order 4, we expect, according to Equation~\ref{EQU:N_sites}, a total of 12 subsites that form 6 magnetic classes, in agreement with our observations. Furthermore, $\mathrm{C_{2v}}$ is a polar point group\textsuperscript{\cite{shea_principles_2005}} that allows for a static electric dipole along the two-fold rotation axis. This may explain the significant spectral diffusion linewidth of tens of megahertz observed for single dopants in the proximity of interfaces.\textsuperscript{\cite{gritsch_optical_2024}} In addition, this symmetry allows the mixture of the $4f$ and $5d$ orbitals\textsuperscript{\cite[A.6]{phenicie_devices_2021}} so that electric dipole transitions between the $4f$ levels are no longer strictly forbidden by parity and may contribute to the radiative decay of the Y$_1\leftrightarrow$Z$_1$ transition. This is consistent with the reported optical lifetime of $\SI{142}{\micro\second}$ for site A,\textsuperscript{\cite{gritsch_narrow_2022}} which is more than ten-fold shorter than the expected magnetic dipole decay in silicon.\textsuperscript{\cite{dodson_magnetic_2012}}

\subsubsection{Site B}
For site B, fitting the spin Hamiltonian is more difficult. It evidently features a lower symmetry which means that more line features have to be extracted in order to fully determine the fit. At the same time, the signal-to-noise ratio is significantly lower, both because the fluorescence is distributed over more spectral features and because site B is overall less bright on the studied sample. Consequently, the peak extraction algorithm as applied for site A and described in section~\ref{SEC:data_preprocessing} could not reliably discriminate between peaks belonging to site B and those of other sites. This precludes a quantitative analysis of the g-tensor of site B.

Nevertheless, we can hypothesize the site symmetry from a qualitative analysis of the recorded spectra. In Table~\ref{TAB:intersections}, we summarize the number of classes of distinguishable emitters for different point groups in two points of high symmetry, when the magnetic field is aligned with the twofold and threefold rotational axis of the crystal, respectively. In Figure~\ref{fig2}~(g), these correspond to field rotation angles of $0^\circ$ and $\arccos \left( \frac{1}{\sqrt{3}} \right)\approx 54.7^\circ$. In particular, at $0^\circ$ and $54.7^\circ$, we observe two, respectively three points of intersection. This rules out all point groups except for $\mathrm{C_s}$ and $\mathrm{C_1}$. Compared to the spin-preserving transitions, more intersections at $0^\circ$ and $54.7^\circ$ are observed on the spin flip transitions in Figure~\ref{fig2}~(h). However, the presence of four crossings at $0^\circ$ is incompatible with the assumption that the crystal is not strained and the magnetic field is applied along a crystal axis, as even a defect with $\mathrm{C_1}$ would produce at most three intersection points in this scenario. This suggests that the additional non-degeneracy in the spin flip transitions is an artifact of straining or slight sample misalignment. When taking this into account, $\mathrm{C_s}$ is most likely the symmetry of site B. 

\begin{table}[]
    \centering
    \begin{tabular}{c c c}
    \toprule
        Point group & $\vec{B} \parallel \langle 100\rangle$ & $\vec{B} \parallel \langle 111\rangle$ \\
        \midrule
         $\mathrm{C_1}$& 3 & 4 \\
         $\mathrm{C_2}$& 3 & 2 \\
         $\mathrm{C_s}$& 2 & 3\\
         $\mathrm{C_3}$& 1 & 2 \\
         $\mathrm{C_{2v}}$& 2 & 2 \\
         \bottomrule
    \end{tabular}
    \caption{Number of classes of distinguishable emitters when a magnetic bias field is applied along directions of high symmetry when the g-tensor respects a certain point symmetry group.}
    \label{TAB:intersections}
\end{table}

This point group of order 2 suggests that there are 24 subsites that are organized in 12 magnetic classes. A single principal axis of each g-tensor is restricted to a $\langle 110\rangle$ axis by symmetry, with the other two axes lying somewhere in the plane orthogonal to it. These latter two axes do not need to coincide for $\mathbf{g}_\mathrm{g}$ and $\mathbf{g}_\mathrm{e}$. As a subgroup of $\mathrm{C_{2v}}$, $\mathrm{C_{s}}$ is likewise polar. Given the lack of a rotational axis, the orientation of the static electric dipole may lie anywhere within the $\{110\}$ mirror plane. As for site A, this symmetry also restricts the possible positions of the erbium dopant which needs to be contained in the $\{110\}$ mirror plane of the site, as illustrated in Figure~\ref{fig3}~(b).

Comparing the spin-flip transitions of site A and B (Figure~\ref{fig1} (f) and (h)), one notes a clear similarity between the two. This suggests that also for site B, $\mathbf{g}_\mathrm{g}$ and $\mathbf{g}_\mathrm{e}$ feature a significant degree of anisotropy, with the biggest eigenvalue belonging to the principal axis along $\langle 110\rangle$ which is imposed by the $\{110\}$ mirror plane of $\mathrm{C_s}$. In this case, we can directly read off the highest principal values from Figure~\ref{fig2}~(c,d) where we apply the field along $(1\Bar{1}0)$, i.e. directly along the simultaneous principal axis of both g-tensors. We identify the highest splitting in (d) with the sum of the g-factors and the peak with the largest thermal population in (c) with the g-factor difference. We also note that $g^\mathrm{(eff)}_\mathrm{g}>g^\mathrm{(eff)}_\mathrm{e}$ based on the relative brightness of the lines in (c). From that, we estimate $g^\mathrm{(1)}_\mathrm{g}=15.7$ and $g^\mathrm{(1)}_\mathrm{e}=13.2$, which is close to the highest principal values found for site A ($g^\mathrm{(1)}_\mathrm{g}=16.6$ and $g^\mathrm{(1)}_\mathrm{e}=13.7$).
Together with the comparable magnitude of their CF splittings and the fact that they favor the same annealing conditions, this could indicate a close relationship between both sites that may be uncovered in future work.

\section{Crystal field Hamiltonian}
Compared to the effective spin Hamiltonian, the CF Hamiltonian offers a more fundamental description of the erbium spectrum. In this approach, one considers a perturbation of the free ion Hamiltonian:
\begin{equation}
    \mathcal{H}=\mathcal{H}_\mathrm{free}+\mathcal{H}_\mathrm{CF}.
\end{equation}
Here, $\mathcal{H}_\mathrm{CF}$ breaks the rotational symmetry of $\mathcal{H}_\mathrm{free}$ and lifts the degeneracy of the J-states. In its most general form, it can capture the mixing of the free ion levels. A reliable fit of this most general CF Hamiltonian therefore requires a rather extensive amount of spectroscopic data for a wide range of wavelengths. For erbium in silicon, acquiring such data may be challenging, if not impossible, due to the large absorption in silicon at shorter wavelengths resulting from its small band gap.

However, the fitting can be largely simplified under the assumption that the crystal field only mixes states within the individual $J$-multiplet of the free ion solution,\textsuperscript{\cite{weber_paramagnetic_1964}} following the historic approach of Stevens.\textsuperscript{\cite{stevens_matrix_1952}} This assumption is justified whenever the $J$-multiplets are sufficiently separated compared to the energy scale of the perturbation. Even then, it may be inadequate to describe for example the electric dipole strength of $4f\leftrightarrow 4f$ transitions, which without the admixture of other orbitals always remains electric-dipole forbidden by parity.

For site A, the $\mathrm{C_{2v}}$ symmetry reduces the number of free parameters sufficiently, such that a fit can be determined with the measured CF splittings presented in Figure~\ref{fig1}~(a) and Table~\ref{TAB:CF}, which also contains the energy levels that could not be determined in.\textsuperscript{\cite{gritsch_narrow_2022}} In the Steven's operator equivalent approach, the CF Hamiltonian for a given $J$-multiplet then takes the form:\textsuperscript{\cite{weber_paramagnetic_1964}}
\begin{equation}
\begin{split}
\mathcal{H}_\mathrm{CF} = B_2^0 O_2^0 + B_2^2 O_2^2 + B_4^0 O_4^0 + B_4^2 O_4^2 + B_4^4 O_4^4 + \\ + B_6^0 O_6^0 + B_6^2 O_6^2 + B_6^4 O_6^4+ B_6^6 O_6^6.
\end{split}
\end{equation} 
Here, $O_l^m=O_l^m(L,S,J)$ are the Steven's operator equivalents expressed in the basis $x^\prime, y^\prime, z^\prime$ imposed by the $\mathrm{C_{2v}}$ symmetry and $B_l^m$ are real-valued CF parameters. For a different multiplet with quantum numbers $\tilde{L},\tilde{S},\tilde{J}$, the corresponding Hamiltonian can then be constructed by forming the corresponding operator equivalents $O_l^m=O_l^m(\tilde{L},\tilde{S},\tilde{J})$ and transforming the CF parameters according to\textsuperscript{\cite[eq.(A5)]{weber_paramagnetic_1964}}

\begin{equation}
\begin{aligned}
\frac{\tilde{B}_l^m}{B_l^m}&=  (-1)^{\tilde{J}-J}\,  \frac{2\tilde{J}+1}{2J+2} \times \\ & \times \sqrt{\frac{(2\tilde{J}-l)! (2J+l+1)!}{(2J-l)! (2\tilde{J}+l+1)!}}\,
\frac{
\begin{Bmatrix}
  \tilde{J} & \tilde{J} & l \\
  L & L & S 
\end{Bmatrix} }{
\begin{Bmatrix}
 J & J & l \\
  L & L & S 
\end{Bmatrix}}
\end{aligned}
\end{equation}

where the expression in the curly brackets are {\it 6-j} symbols. This means that we may predict the 7+6 CF splittings of \gstate and \estate with a given set of nine $B_l^m$ parameters. This allows us to perform a determined fit with limited experimental data and at little computational cost. Previously, this approach has been successfully applied to describe several erbium implantation centers,\textsuperscript{\cite{hughes_optically_2019}} as well as other rare earths\textsuperscript{\cite{hughes_crystal_2014}} in silicon.

\begin{table*}[]
        \centering
        \begin{tabular}{c c c c c c c c}
        \toprule
        
        \multicolumn{3}{c}{Ground state multiplet \gstate} & \multicolumn{3}{c}{Excited state multiplet \estate} & \multicolumn{2}{c}{Fitted CF parameters}
        \\\cmidrule(lr){1-3}\cmidrule(lr){4-6}\cmidrule(lr){7-8}

           Level & measured [GHz] & predicted [GHz] &Level & measured [GHz] & predicted [GHz] & Parameter  & [GHz] \\
           \cmidrule(lr){1-3}\cmidrule(lr){4-6}\cmidrule(lr){7-8}
            Z$_1$ & 0.0 & 0.0& Y$_1$ & 0.0 & 0.0& $B_2^0$ & $3.6\times10^{1}$\\ 
            Z$_2$ & 2634.2 & 2610.5& Y$_2$ & 2417.9 & 2451.9& $B_2^2$ & $2.6\times10^{1}$\\ 
            Z$_3$ & 3095.9 & 3079.4& Y$_3$ & 3080.9 & 3104.8 & $B_4^0$ & $-1.0\times10^{-2}$\\ 
            Z$_4$ & 5388.9 & 5415.7& Y$_4$ & 3685.3 & 3639.4 & $B_4^2$ & $-7.5\times10^{-3}$\\
            Z$_5$ & 6414.3 & 6414.7& Y$_5$ & 4057.0 & 4008.2 & $B_4^4$ & $-1.2\times10^{-1}$\\ 
            Z$_6$ & 7477.9 & 7504.7& Y$_6$ & 6175.9 & 6241.3 & $B_6^0$ & $-6.2\times10^{-4}$\\ 
            Z$_7$ & 9456.0 & 9425.6& Y$_7$ & 7832.6 & 7846.7 & $B_6^2$ & $2.1\times10^{-2}$\\ 
            Z$_8$ & 12264.0 & 12242.6&  & & & $B_6^4$ & $4.2\times10^{-3}$\\ 
              &  &  &  & & & $B_6^6$ & $1.7\times10^{2}$\\
        \bottomrule
        \end{tabular}
        \caption{Energy splitting as measured and predicted by the CF Hamiltonian fit for the \gstate (left) and \estate (center) multiplet. Right: Fitted crystal field parameters.} 
     
    \label{TAB:CF}
\end{table*}

The results of the fit, both in terms of CF parameters and calculated CF splittings, are shown in Table~\ref{TAB:CF}. We reproduce the splittings with an accuracy of about \SI{30}{\giga\hertz} (RMSD), which is small compared to the overall splitting of the ground state multiplet of about $\SI{12}{\tera\hertz}$. 
The remaining deviation is larger than the statistical error of the fit owing to the limitation of the used CF model that does not consider higher-lying orbitals. Details regarding the fitting procedure can be found in section~\ref{SEC:CFFitting}.

As a consistency check, we derive the Zeeman interaction from the crystal field fit and compare it to the prediction of the spin Hamiltonian that has been derived. The principal values of the g-tensor can be calculated by evaluating: 
\begin{equation}
\begin{split}
g^{(1)}&=2i\,g_J\langle \uparrow |J_{y^\prime}|\downarrow\rangle   \,\, \\
g^{(2)}&=2\,g_J\langle \uparrow|J_{x^\prime}|\downarrow\rangle \, \,\, \\ g^{(3)}&=2\,g_J\langle \uparrow |J_{z^\prime}| \uparrow\rangle.
\end{split}
\end{equation}
Here, $\ket{\uparrow}$ and $\ket{\downarrow}$ are the two eigenstates of the CF Hamiltonian corresponding to the lowest energy Kramers doublet in the respective $J$-manifold. $g_J$ is the Landé factor for the respective $J$-multiplet ($g_J=6/5$ and $g_{\tilde{J}}=72/65$ for \gstate and \estate respectively) and $J_{x^\prime}, J_{y^\prime}, J_{z^\prime}$ are the corresponding spin matrices.\textsuperscript{\cite{hughes_optically_2019}} Comparing the predictions of the CF Hamiltonian with the experimentally determined g-tensor values of the spin Hamiltonian, we find
\begin{table}[H]
        \centering
{\centering
\begin{tabular}{lcccccc}
\toprule
     & $g_\mathrm{g}^{(1)}$ & $g_\mathrm{g}^{(2)}$ & $g_\mathrm{g}^{(3)}$ & $g_\mathrm{e}^{(1)}$ & $g_\mathrm{e}^{(2)}$ & $g_\mathrm{e}^{(3)}$\\
     \midrule
     Spin Hamiltonian& 16.6 & 0.4&0.6&13.7&0.2&0.2\\
     Crystal field Hamiltonian& 16.5 &  1.4 &   0.8 & 13.9&  0.2&   0.0  \\
     \bottomrule
\end{tabular}}
\end{table}

The principal axes are fully imposed by symmetry and thus identical for the two Hamiltonians. Both models predict the same high anisotropy of the g-tensors and we find a comparable principal value along this direction.

\section{Summary and Outlook}
In summary, we have performed a detailed analysis of the magnetic interaction for two recently discovered crystallographic sites of erbium dopants in silicon. For site A, we have fully reconstructed the spin Hamiltonian and also derived its point symmetry $\mathrm{C_{2v}}$. This allowed us to perform a partial fitting of the crystal field, restricted to the splittings of the \gstate and \estate manifold. Comparing the predictions of the corresponding crystal field Hamiltonian with the predictions of the spin Hamiltonian, we find a good agreement.

With the CF Hamiltonian, one can now describe the magnetic interaction of this site on a more fundamental level and predict the Zeeman effect of all Kramers doublets in both the \gstate and \estate multiplets. Furthermore, moving beyond the effective spin Hamiltonian approach, one can predict avoided crossings and the corresponding non-linearities that arise within the two multiplets when higher magnetic fields are applied. This may give rise to transitions that are significantly less sensitive to fluctuations of the magnetic environment.

For site B, a quantitative fitting turned out more difficult given the signal-to-noise ratio of the data. Still, a qualitative analysis suggests $\mathrm{C_s}$ for its point symmetry. As this is a subgroup of $\mathrm{C_\mathrm{2v}}$, site B is expected to share certain properties with site A, including a permanent electric dipole and electric-dipole-allowed $4f\leftrightarrow 4f$ transitions.

Beyond its immediate use for predicting spectroscopic properties, determining the symmetry properties of these two sites is a first step towards understanding their microscopic structure. This in turn may allow improving the yield of doping procedures or even enable purposeful site engineering. As such, it is a crucial step towards high-performing erbium-based silicon devices for distributed quantum information processing in this emerging hardware platform.

\section{Methods}

\subsection{Experimental setup and data acquisition} \label{subsec:setup}

The sample is fabricated from a float-zone-grown silicon-on-insulator wafer with a device layer thickness of $\approx \SI{2}{\micro\meter}$, similar to the device used in.\textsuperscript{\cite{gritsch_narrow_2022}} Erbium dopants are implanted without isotopic selectivity under an angle of $\SI{7}{\degree}$ at the Helmholtz-Zentrum Dresden-Rossendorf. To achieve an approximately homogeneous dopant concentration of $\SI{1e16}{\per\centi\meter\cubed}$, we use three implantation runs with implantation energies of $\SI{1.5}{\mega\electronvolt}$, $\SI{2.5}{\mega\electronvolt}$ and $\SI{4}{\mega\electronvolt}$ and adapted doses of $\SI{4e11}{\centi\meter^{-2}}$, $\SI{6e11}{\centi\meter^{-2}}$, and $\SI{1e12}{\centi\meter^{-2}}$, respectively. During the implantation, the sample is kept at ambient temperature. To study the erbium dopants integrated at site A, we perform a post-implantation annealing step at $\SI{800}{\kelvin}$ with a hold time of $\SI{1}{\min}$ in a rapid thermal annealing oven. After implantation, $\SI{8}{\milli\meter}$-long nanophotonic rib waveguides are fabricated using electron beam lithography and reactive ion etching at cryogenic temperatures using a fluorine chemistry. Then, a standard single mode fiber (SMF-28) is glued (Norland NOA 88) to the facet of a waveguide. This approach allows rotating the sample in the field without requiring re-alignment of the fiber-chip coupling. However, it only gives a moderate one-way coupling efficiency of $\approx \SI{4}{\percent}$, limited by the mode overlap between the fiber and the waveguide.

The device is mounted in a closed-cycle Helium cryostat (Kiutra S-type Optical). To avoid spin polarization, the measurements are performed at a temperature of $\approx \SI{10}{\kelvin}$ using a resistive heater on the sample stage. The cryostat is equipped with a superconducting solenoid magnet which can apply fields up to $\SI{3}{\tesla}$. To facilitate the required precise rotation around the [110] axis, the sample is mounted on a cryogenic rotation stage (Attocube ANRv51/RES/LT) with resistive encoder that gives a relative angular accuracy below $\SI{0.1}{\degree}$. The mount is designed so that one can rotate the sample with fibers glued to its facets.

We then perform pulsed, resonant spectroscopy. For this purpose, a series of acousto-optic modulators is used to generate $\SI{100}{\micro\second}$ long pulses from a cw-laser (Toptica CTL 1500). We sweep the excitation laser frequency in a range of $\SI{89}{\giga\hertz}$ ($\SI{101}{\giga\hertz}$) and $\SI{\pm 63}{\giga\hertz}$ ($\SI{\pm 127}{\giga\hertz}$) for the spin-preserving and spin-flip transitions of site A (site B), respectively, in steps of $\approx \SI{125}{\mega\hertz}$. The excitation wavelength is determined with a HighFinesse WS8-10 wavemeter with a precision of $\SI{8}{\mega\hertz}$. After each laser pulse, the fluorescence is detected using a superconducting nanowire single-photon detector (IDQuantique, dark count rate: $<\SI{150}{\hertz}$). To improve the signal-to-noise ratio that is impeded by the off-resonant background observed in erbium-doped silicon,\textsuperscript{\cite{gritsch_narrow_2022}} we use a free-space long-pass filter (Semrock BLP01-1550R), which transmits light emitted into the high-lying CF levels of site A and B, but blocks resonant fluorescence and off-resonant emission from erbium dopants with smaller CF splittings.

\subsection{Data preprocessing}\label{SEC:data_preprocessing}
In order to fit our theoretical model to the experimental data, we need to extract the field-dependent positions of the different Zeeman transitions in the recorded spectra. A major challenge here is the presence of fluorescent features that originate from other erbium sites. Thus, we employ different strategies for extracting the position of the spin-flip and spin-preserving transitions.

\begin{figure*} 
     \includegraphics[width=1\linewidth]{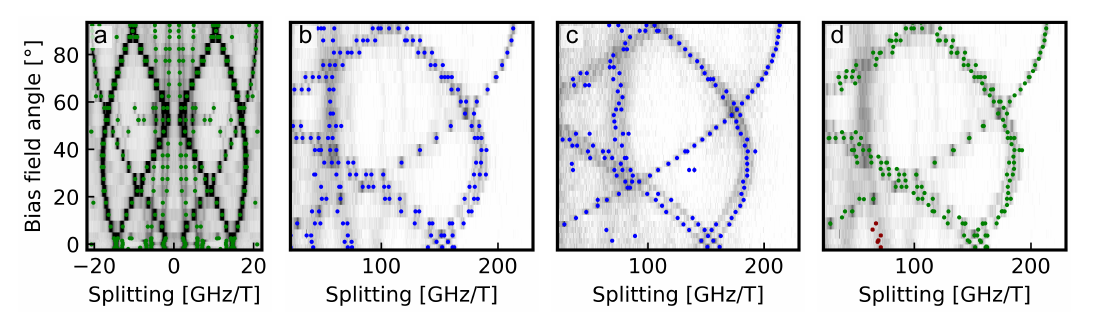}
        \caption{(a) Symmetrized fluorescence map and identified peaks for the spin-preserving transitions of site A. (b,c) Fluorescence map and identified peaks for the spin-flip transitions of site A at \SI{350}{\milli\tesla} (b) and \SI{250}{\milli\tesla} (c) bias field. (d) Peaks that have been identified to be common between the data sets in (b) and (c). Five peaks (red crosses) belong to a different site such that they have to be discarded. However, the automated peak extraction algorithm fails as there is some overlap between the higher splitting branch in panel (b) and the lower branch in panel (c). Thus, these points have been removed manually for the subsequent fitting procedure. }
        \label{FIG:SFprocess}
\end{figure*}

The latter are relatively bright, and their complete symmetric spectrum, i.e. both the $f_\mathrm{+-}$ and the $f_\mathrm{-+}$ branches, could be recorded without overlapping with bright fluorescent features from other sites. To also discard weak features from other sites in the peak identification procedure, we mirror the spectrum around the zero-field emission wavelength of site A and add it to its non-mirrored self, thus increasing the signal in relation to the background peaks. With this, the positions of the fluorescence peaks can be extracted reliably using an algorithm that is conditioned on peak amplitude and prominence. The result of the peak extraction is shown in Figure~\ref{FIG:SFprocess}(a).

The preprocessing for the spin-flip transitions is more complicated, as they are much dimmer due to the small branching ratio of only a few percent. In addition, overlap with other fluorescent features of much stronger prominence has hindered recording a symmetric spectrum. Furthermore, even for the cleaner side of the spectrum, overlapping features have been observed at all bias fields.

Therefore, we record two datasets, shown in Figure~\ref{FIG:SFprocess}~(b,c), at different magnetic fields, and scale them such that the x-axis corresponds to the peak splitting from the zero-field wavelength. In this way, one expects that peaks that belong to site A end up at the same position, while peaks from other sites would appear at different positions in the spectra. Thus, after peak extraction conditioned on amplitude and prominence, peaks from other sites can be discarded by keeping only those that appear in both data sets. The coinciding peaks are shown in Figure~\ref{FIG:SFprocess}(d).

\subsection{Spin Hamiltonian fitting}\label{SEC:Fitting}
As motivated in section~\ref{SEC:spinHamil}, the spin Hamiltonians of the $Z_1$ and $Y_1$ doublets are fully determined by the g-tensors $\mathbf{g}_\mathrm{g}$ and $\mathbf{g}_\mathrm{e}$, respectively. As symmetric $3\times3$ tensors, we parameterize them with six parameters each
\begin{equation}
    \mathbf{g}_\mathrm{g}=\begin{pmatrix}
  p_1 & p_4 & p_5\\
  p_4 & p_2 & p_6\\
  p_5 & p_6 & p_3\\
\end{pmatrix},~
  \mathbf{g}_\mathrm{e}=\begin{pmatrix}
  p_7 & p_{10} & p_{11}\\
  p_{10} & p_8 & p_{12}\\
  p_{11} & p_{12} & p_9\\
\end{pmatrix}.
\end{equation}
Then, we calculate the frequencies of the four different Zeeman transitions according to Equation~\ref{EQU:branches}. In order to predict the Zeeman transition frequencies for all magnetic classes, we make use of Equation~\ref{EQU:syms} to generate 24 pairs of g-tensors. In general, the model will therefore predict $4 \times 24 = 96$ transitions for any given field orientation.

For the loss function, we sum over the relative distance squared of all data points to the closest model prediction
\begin{equation}
\begin{aligned}
    &\mathcal{L}(p_1,\dots,p_{12}) = \\ 
    &\sum_i \min_j \left[\left(\frac{ f_i-m_j\left(\vec{B}_i,p_1,\dots,p_{12}\right)}{f_i}\right)^2~\right].
\end{aligned}
\end{equation}
Here, $f_i$ are the measured peak positions, $\vec{B}_i$ the field at which they were measured and $m_j\left(\vec{B}_i,p_1,\dots,p_{12}\right)$ are all model predictions for the given field and parameter set. The form of this loss function allows to fit the data without making prior assumptions regarding which data point is associated with which line or whether the dataset is complete, i.e. whether for a given field orientation all transitions have been identified.

A drawback of this approach is that it does not disincentivize the prediction of transition lines which are not supported by the measured data. The fit may thus tend towards results with too low symmetry while using the additional, non-physical degrees of freedom to minimize the loss originating from single outliers. We overcome this problem by adding an additional term to the loss function. For each model line, we add the relative distance squared of the ten closest data points. In this way, we still don't assume completeness of the data, but require a minimum of empirical support for each feature of the model. In order to ensure that data for all Zeeman transitions are present, the peaks of the spin-flip transitions, shown in Figure~\ref{FIG:SFprocess}~(d) are mirrored around the zero field frequency before fitting. Their weight is halved compared to the data of the spin-preserving lines in Figure~\ref{FIG:SFprocess}~(a) such that they do not contribute to the fit excessively.

\begin{figure*} 
     \includegraphics[width=1\linewidth]{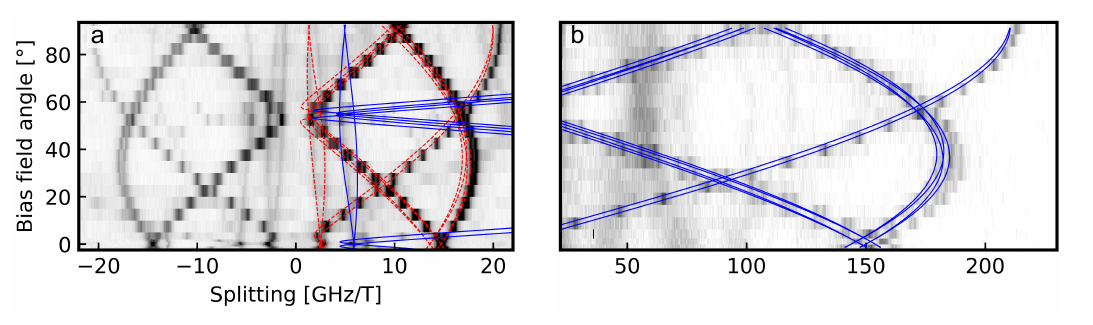}
        \caption{Spin-preserving transitions (a) and spin-flip transitions (b) of site A, with the result of the unrestricted fit superimposed. Model prediction of the spin-preserving transition are shown in red, spin-flip transition are shown in blue.}
        \label{FIG:genfit}
\end{figure*}

The loss function in this form is quite multimodal. In order to ensure that a global maximum is found, we use a basin hopping algorithm for minimization. Given an appropriate initial guess, the algorithm is observed to converge to very similar parameter values in different runs. Beyond the existence of several local minima, the loss function is expected to also feature several global maxima. These, however, correspond to different g-tensors that are related by Equation~\ref{EQU:syms} such that we may expect that there exists a global minimum that is unique up to a crystal symmetry transformation. We show the fitting result without imposed symmetry, corresponding to the g-tensor from Equation~\ref{EQU:genfit}, in Figure~\ref{FIG:genfit}.

For the fit with enforced symmetry, we make use of the fact that the g-tensor has to feature a principal axis along every axis of rotation, and orthogonal to every plane of symmetry.\textsuperscript{\cite{spaeth_structural_2012}} For $\mathrm{C_{2v}}$, this restricts one principal axis to $\langle 001\rangle$, along the two-fold rotation axis, another to $\langle 110\rangle$, orthogonal to the mirror plane, and the third to $\langle 1\Bar{1}0\rangle$, orthogonal to the two other axes. As all principal axes are determined by symmetry, each g-tensor can be parameterized by its three eigenvalues, so
\begin{equation}
\begin{split}
    \mathbf{g}_\mathrm{g}&=\mathcal{R}_\mathrm{z}^{-1}(\pi/2)\cdot\mathcal{D}(p_1,p_2,p_3)\cdot\mathcal{R}_\mathrm{z}(\pi/2) \\
    \mathbf{g}_\mathrm{e}&=\mathcal{R}_\mathrm{z}^{-1}(\pi/2)\cdot\mathcal{D}(p_4,p_5,p_6)\cdot\mathcal{R}_\mathrm{z}(\pi/2) \\
\end{split}
\end{equation}
where $\mathcal{D}$ are diagonal matrices in the crystal axis basis and $\mathcal{R}_\mathrm{z}$ are matrices for rotation around the $z$-axis. In order to fit the model, we proceed as for the fit without imposed symmetry.

\subsection{Crystal field Hamiltonian fitting}\label{SEC:CFFitting}
Although the optimization problem is overdetermined since we are looking for nine $B_l^m$ parameters from the 7+6 measured energy differences for the ground and excited states, it is important to verify its convergence by varying the initial parameters. In order to initiate the optimization algorithm (Powell's method), we propose to start from a known situation well adapted to silicon, the cubic symmetry, as an initial guess for the algorithm. The crystal
field is then restricted to four non-zero terms, including only two independent parameters $B_4^0$ and $B_6^0$ and imposing $B_4^4=5B_4^0$ and $B_6^4=-21B_6^0$.\textsuperscript{\cite{lea_raising_1962}}
Our search strategy is to vary the initial values of $B_4^0$ and $B_6^0$ for a cubic symmetry, and for each pair let the algorithm converge to a set of nine non-zero optimal parameters corresponding to the $\mathrm{C_\mathrm{2v}}$ symmetry by minimizing the RMSD between the calculated and experimental values expressed in GHz as summarized in Table \ref{TAB:CF}. By varying the initial parameters, we finally select only the best solution that minimizes the overall RMSD, yielding the solution in Table \ref{TAB:CF}.

\medskip
\textbf{Acknowledgements} \par
We thank Kilian Baumann for his technical contribution at an early stage of the project. This project received funding from the Deutsche Forschungsgemeinschaft (DFG, German Research Foundation) under Germany's Excellence Strategy - EXC-2111 - 390814868, from the German Federal Ministry of Education and Research (BMBF) via 16KISQ046, and from the Munich Quantum Valley that is funded by the Bavarian state government via the Hightech Agenda Bayern Plus.

\medskip

\bibliography{bibliography.bib}

\end{document}